\documentclass[%
 reprint,
 amsmath,amssymb,
 aps
]{revtex4-2}
\usepackage{subfigure}
\usepackage{braket}
\usepackage{lipsum} 
 \usepackage[colorlinks=true]{hyperref}
\usepackage{orcidlink}
\usepackage{graphicx}
\usepackage{dcolumn}
\usepackage{bm}\allowdisplaybreaks
\makeatletter
\newenvironment{mywidetext@grid}{%
  \par\ignorespaces
  \setbox\mywidetext@top\vbox{%
   \vskip15\p@
   \hb@xt@\hsize{%
    \leaders\hrule\hfil
    \vrule\@height6\p@
   }%
   \vskip6\p@
  }%
  \setbox\mywidetext@bot\hb@xt@\hsize{%
    \vrule\@depth6\p@
    \leaders\hrule\hfil
  }%
  \onecolumngrid
  \dimen@\ht\mywidetext@top\advance\dimen@\dp\mywidetext@top
  \cleaders\box\mywidetext@top\vskip\dimen@
  \let\set@footnotewidth\set@footnotewidth@ii
}{%
  \par
  \setbox\mywidetext@bot\vbox{%
   \hb@xt@\hsize{
   }%
   \vskip14\p@
  }%
  \dimen@\ht\mywidetext@bot\advance\dimen@\dp\mywidetext@bot
  \cleaders\box\mywidetext@bot\vskip\dimen@
  \twocolumngrid\global\@ignoretrue
  \@endpetrue
}%
\newbox\mywidetext@top
\newbox\mywidetext@bot
\appdef\class@documenthook{%
 \twocolumn@sw{%
  \let@environment{mywidetext}{mywidetext@grid}%
  \let\title@column\title@column@grid
  \let\close@column\close@column@grid
 }{%
  \let@environment{mywidetext}{mywidetext@galley}%
  \preprintsty@sw{%
  }{%
   \galley@sw{%
    \let\mywidetext@outdent\galley@outdent
   }{%
   }%
  }%
 }%
}%
\appdef\class@inithook{%
 \@ifxundefined\title@column{%
  \let\title@column\title@column@default
 }{}%
}%
\makeatother
\begin{document}

\title{Complex fixed points of the non-Hermitian Kondo model in a Luttinger liquid}

\author{SangEun Han\,\orcidlink{0000-0003-3141-1964}}
\affiliation{Department of Physics, University of Toronto, Toronto, Ontario M5S 1A7, Canada}
\author{Daniel J.~Schultz\,\orcidlink{0000-0003-0567-850X}}
\affiliation{Department of Physics, University of Toronto, Toronto, Ontario M5S 1A7, Canada}
\author{Yong Baek Kim}
\affiliation{Department of Physics, University of Toronto, Toronto, Ontario M5S 1A7, Canada}

\date{\today}

\begin{abstract}
Non-Hermitian physics in open quantum many-body systems provides novel opportunities for discovery of exotic quantum phenomena unexpected in Hermitian systems. A previous study of the non-Hermitian Kondo problem in ultra-cold atoms reports reversion of renormalization group flows which violates the $g$-theorem and produces an unusual quantum phase transition. In this work, we study the effect of electron-electron interactions by considering the non-Hermitian Kondo problem in a Luttinger liquid. 
By performing a perturbative renormalization group analysis to two-loop order, we find that the interplay between non-Hermitian Kondo couplings and electron-electron interactions can produce a pair of complex fixed points. Complex fixed points have often been discussed in an attempt to understand extremely long correlation length of Hermitian systems with weakly first-order transitions. Here, we show that complex fixed points arise naturally and can be physically realized in open quantum systems. We discuss consequences of the complex fixed points and future directions.
\end{abstract}

\maketitle

\section{Introduction}

The recent popularity of ultracold atomic systems as a playground to engineer exotic quantum many-body physics has presented new opportunities to study open quantum systems \cite{Daley2014,Kuhr2016,Ashida2020,Syassen2008,Barontini2013,Peise2015,Luschen2017}. 
In contrast to typical quantum statistical mechanics, where the system is in equilibrium with an external bath, an open quantum system may lose energy or particles to the bath via dissipation. These dissipative effects manifest as non-Hermitian terms in the effective Hamiltonian of the system \cite{Daley2014,Ashida2020}, and allow phenomena prohibited by the unitarity condition of ordinary Hermitian quantum theory, for example continuous phase transitions without gap-closing \cite{Matsumoto2020} and anomalous enhancement of the superfluid correlation with spontaneous $\mathcal{PT}$-symmetry breaking \cite{Ashida2017}. 

The most common way to design experiments exhibiting non-Hermitian physics is through ultracold atoms. In recent years, experimentalists have been able to construct multi-band models, which opened up the possibility for systems such as the Anderson or Kondo models \cite{White2020,Riegger2018,Gorshkov2010,KanaszNagy2018,Zhang2020}. In particular, Alkaline-earth atoms can have both long-lived metastable excited states playing the role of a localized impurity, while their ground states play the role of conduction electrons \cite{Riegger2018,Gorshkov2010,KanaszNagy2018,Zhang2020}. If we consider two-body loss/gain induced by the inelastic scattering between ground states and excited states \cite{Riegger2018,Scazza2014,Hofer2015,Pagano2015}, it introduces the non-Hermitian Kondo effect \cite{Nakagawa2018}. Previous studies of the non-Hermitian Kondo problem using a renormalization group (RG) analysis have revealed a reversion of the RG flow; even though the initial parameters are in the antiferromagnetic Kondo coupling regime, the system can flow toward to the non-Kondo phase characterized by the Gaussian fixed point \cite{Nakagawa2018}. This result, which violates the $g$ theorem \cite{Affleck1991,Friedan2004}, also has been confirmed by the Bethe ansatz. These previous studies however have neglected the electron-electron interaction. 

In this work, we analyze the non-Hermitian Kondo effect in a Luttinger liquid to investigate the effect of electron-electron interactions in non-Hermitian systems. We consider a one-dimensional model of interacting electrons with non-Hermitian forward and backward Kondo interactions with a spin-1/2 localized impurity, the non-Hermitian part of which arises from the interactions with the environment. We analyze the problem by computing the RG flow equations up to two-loop order. 
In the case of the Hermitian Kondo interactions, the system flows toward the isotropic strong coupling limit with Kondo singlet \cite{furusaki1994,per1995,per1996}.
In the case of the non-Hermitian Kondo interactions without electron-electron interactions, there is the previously observed reversion of the RG flow near the Gaussian fixed point \cite{Nakagawa2018,Lourenco2018}. However, introducing a weak repulsive electron-electron interaction yields a pair of complex fixed points on the real-forward and imaginary-backward Kondo coupling plane. The fixed point values depend on the strength of the electron-electron interaction, and have a complex-valued scaling dimension. These fixed points are stable in the isotropic limit.
These complex fixed points are in the perturbative regime which is controlled by the strength of the electron-electron interaction and is the result of the interplay between the non-Hermitian Kondo interaction and electron-electron interaction, and is absent when only one of the two is present.
Interestingly, due to the imaginary part of the scaling dimension, the RG flow shows spiral-like behaviors near the complex fixed points \cite{Calabrese2002,Calabrese2003,KHMELNITSKII197859,Dorogovtsev1980,Yerzhakov2018,Yerzhakov2021}.
In the anisotropic case (XXZ Kondo interaction), the complex fixed point is no longer stable, but in its vicinity the RG flow shows walking behavior \cite{Piai2010,Gorbenko2018a,Gorbenko2018b,ma2019,ma2020} and hence spends a long time there, leading to an unusually long correlation length. Eventually however, the RG runs to the isotropic (antiferromagnetic) strong coupling limit or an Ising fixed plane with strong backward-scattering Kondo coupling. 

This result provides a novel example of the interplay between the electron-electron interaction and non-Hermitian interaction. Such complex fixed points have been mainly investigated in a theoretical setting where a Hermitian system with weakly first-order transitions is analytically continued to the complex plane \cite{Gorbenko2018a,Gorbenko2018b,ma2019,ma2020}. The presence of complex fixed points close to the real axis in the complex plane was suggested to be responsible for the weakly first-order behavior and the long correlation length in the Hermitian system \cite{Gorbenko2018a,Gorbenko2018b,ma2019,ma2020}. In the current work, we show that complex fixed points can be realized in open quantum systems via electron-electron interaction. Our work has direct implications to experiments on open quantum systems and provides an exciting platform for making connections between complex conformal field theory and cold-atom experiments.

The remainder of the paper is organized as follows: In Sec.~\ref{sec:model}, we introduce the Hermitian Kondo model and non-Hermitian Kondo interactions in a Luttinger liquid. We consider two cases: one where the Kondo interactions are isotropic, and also XXZ Kondo interactions. In Sec.~\ref{sec:rganalysis}, we perform the renormalization group analysis for the isotropic and anisotropic models. 
First, we review the results of the isotropic Hermitian case and analyze the non-Hermitian case. We show how the complex fixed point arises in the simultaneous presence of electron-electron interactions and non-Hermitian Kondo interactions, and discuss its critical properties. After that, we generalize the model to the anisotropic case and discuss the general behavior and the walking behavior near the complex fixed point. In Sec.~\ref{sec:conclusion}, we summarize the results and discuss the implications of our work.

\section{Model}\label{sec:model}
In this work, we start by considering the one-dimensional Luttinger liquid,
\begin{align}
H_{\text{LL}}={}&v_{F}\sum_{k,\sigma}k(\psi_{R\sigma,k}^{\dagger}\psi_{R\sigma,k}-\psi_{L\sigma,k}^{\dagger}\psi_{L\sigma,k})\notag\\
&+\frac{g_{2}}{\mathcal{L}}\sum_{k_{1},k_{2}}\sum_{p}(\psi_{L,k_{1}}^{\dagger}\sigma^{0}\psi_{L,k_{1}-p})(\psi_{R,k_{2}}^{\dagger}\sigma^{0}\psi_{L,k_{2}+p}),\label{eq:luttinger_fermion}
\end{align}
where we assume that we are away from half-filling and are on the Tomonaga-Luttinger fixed line to eliminate both umklapp and backward scatterings \cite{Solyom1979,Giamarchi2003,fradkin2013}. $\mathcal{L}$ is the system size, and a one-dimensional spin-1/2 fermion, $\Psi_{\sigma}$ ($\sigma=\uparrow,\downarrow$), is split into left and right movers near the Fermi surface via $\Psi_{\sigma}(x)\approx e^{-ik_{F}x}\psi_{L\sigma}(x)+e^{ik_{F}x}\psi_{R\sigma}(x)$. The left and right movers behave as two different bands.

The electron-electron interaction $g_{2}$ is a repulsive forward inter-band scattering, $g_{2}>0$.
To establish notation for the Luttinger parameters, which will become important later in this work, we bosonize the fermion Hamiltonian Eq.~\eqref{eq:luttinger_fermion}. It can then be expressed as
\begin{align}
H_{\text{LL}}={}&\int\frac{dx}{2\pi}\sum_{a=s,c}\frac{v_{a}}{2}\left(\frac{1}{K_{a}}\Pi_{a}^{2}+K_{a}(\partial_{x}\phi_{a})^{2}\right).
\end{align}
Here, the spin and charge degrees of freedom separate and accordingly $\phi_{s/c}$ are the spin/charge boson fields and $\Pi_{s/c}$ are the corresponding conjugate fields. $v_{s}$ and $v_{c}$ are the velocities of the spin and charge degrees of freedom which are defined by $v_{s}=v_{F}$ and $v_{c}=v_{F}\sqrt{1-(g_{2}/\pi v_{F})^{2}}$, $K_{s}$ and $K_{c}$ are the Luttinger parameters for the spin and charge. For repulsive $g_2\geq 0$, the Luttinger parameters satisfy $K_{c}=\sqrt{\frac{1-g_{2}/\pi v_{F}}{1+g_{2}/\pi v_{F}}}\leq1$, and $K_{s}=1$, which describes the zero temperature scaling behaviors of the Luttinger liquid. 

Returning to the fermionic representation, we add the Kondo interaction between conduction electrons and a spin-1/2 impurity spin $\vec{S}$ localized at the origin ($x=0$), 
\begin{align}
    H_{K}={}&\frac{J_{\perp}}{2}\sum_{i=x,y}(\Psi_{0}^{\dagger}\tau^{i}\Psi_{0})S^{i}+\frac{J_{z}}{2}(\Psi_{0}^{\dagger}\tau^{z}\Psi_{0})S^{z}\notag\\
        \approx{}&
        \frac{J_{\perp}}{2}\sum_{i=x,y}[(\psi_{L,0}^{\dagger}+\psi_{R,0}^{\dagger})\tau^{i}(\psi_{L,0}+\psi_{R,0})]S^{i}\notag\\
        &+\frac{J_{z}}{2}[(\psi_{L,0}^{\dagger}+\psi_{R,0}^{\dagger})\tau^{z}(\psi_{L,0}+\psi_{R,0})]S^{z},
\end{align} 
where $J_{\perp},J_{z}$ represent the strength of the Kondo interaction, $\tau^{i}$ is the Pauli matrix acting on the fermion spin, $\vec{S}$ is a localized impurity spin located at $x=0$, $\Psi_{0}$ and $\psi_{(L,R),0}$ stand for fermion fields located at $x=0$. The Kondo interaction can be split up into forward and backward impurity scattering events $H_{K}=H_{F}+H_{B}$ \cite{furusaki1994} in the following way:
\begin{align}
H_{F}=&\frac{J_{\perp F}}{2}\sum_{i=x,y}[(\Psi_{L,0}^{\dagger}\tau^{i}\Psi_{L,0})+(\Psi_{R,0}^{\dagger}\tau^{i}\Psi_{R,0})]S^{i}\notag\\
&+\frac{J_{z F}}{2}[(\Psi_{L,0}^{\dagger}\tau^{z}\Psi_{L,0})+(\Psi_{R,0}^{\dagger}\tau^{z}\Psi_{R,0})]S^{z},\label{eq:HFaniso}\\
H_{B}=&\frac{J_{\perp B}}{2}\sum_{i=x,y}[(\Psi_{L,0}^{\dagger}\tau^{i}\Psi_{R,0})+(\Psi_{R,0}^{\dagger}\tau^{i}\Psi_{L,0})]S^{i}\notag\\
&+\frac{J_{z B}}{2}[(\Psi_{L,0}^{\dagger}\tau^{z}\Psi_{R,0})+(\Psi_{R,0}^{\dagger}\tau^{z}\Psi_{L,0})]S^{z}.\label{eq:HBaniso}
\end{align}
In contrast to the Kondo model of higher dimensions, which is a theory containing either only left or only right movers, in one dimension both are present. Hence, we have both forward and backward Kondo interactions.
If we take the isotropic limit, $J_{\perp F}=J_{z F}=J_{F}$ and $J_{\perp B}=J_{z B}=J_{B}$, $H_{F}$ and $H_{B}$ have SU(2) symmetry. Note that, in order to later perform the RG analysis, we construct the impurity in terms of a pseudofermion $f_\alpha$ ($\alpha=\uparrow,\downarrow$), whereby $\vec{S}=\frac{1}{2}\sum_{\alpha\beta=\uparrow,\downarrow}f_{\alpha}^{\dagger}\vec{\sigma}_{\alpha\beta}f_{\beta}$. To stay within the physical (singly occupied) subspace, we enforce the constraint $\sum_{\alpha=\uparrow,\downarrow}f_{\alpha}^{\dagger}f_{\alpha}=1$ via introducing a chemical potential term $\lambda\sum_{\alpha} f^\dagger_{\alpha}f_{\alpha}$, which is sent to $\lambda\to\infty$ at the end of the calculation \cite{zhu2002,han2022}. 

So far, we constructed a Hermitian Hamiltonian, $H=H_{\text{LL}}+H_{F}+H_{B}$ with real-valued $J_{\perp F/z F}$ and $J_{\perp B/z B}$. Non-Hermitian terms arise when the coupling to the bath is taken into account via the Lindblad equation \cite{Lindblad1976,Gorini1976,Daley2014,Ashida2020}. The effect of this coupling is described by the jump operators $L_i$, which either add or remove two particles from the system: $L_{i}\propto\Psi_{\uparrow,0}f_{\uparrow},\Psi_{\downarrow,0}f_{\downarrow},\Psi_{\uparrow,0}f_{\downarrow}\pm\Psi_{\downarrow,0}f_{\uparrow}$ \cite{Nakagawa2018} (see Appendix~\ref{app:NH_Kondo}). The jump operators enter into the effective Hamiltonian via the anti-Hermitian term $H_{\text{AH}}=-\frac{i}{2}\sum_{i}L_{i}^{\dagger}L_{i}$, and yield the effective Hamiltonian $H_{\text{eff}}=H+H_{\text{AH}}$. Including $H_{\text{AH}}$ extends the parameter space of the initially real Kondo couplings $J_{\perp F/z F}$ and $J_{\perp B/z B}$ to complex numbers. This leaves us with 8 independent real interaction parameters describing the model.
There are also operators which describe the quantum-jump process of the impurity but we project out such a process to prevent loss of the impurity \cite{Nakagawa2018,Ashida2016,Ashida2017,Ashida2018,Ashida2020,Yamamoto2019,Matsumoto2020}.
We note that, although $H_{\text{eff}}$ technically includes a potential scattering proportional to $\Psi^\dagger_0\sigma^0\Psi_0$, {we ignore it because it does not qualitatively change the results.}

\section{RG analysis}\label{sec:rganalysis}
In this section, we will start by reviewing the results for the Hermitian Kondo effect in a Luttinger liquid. Then, we will explain how this changes when the Kondo interaction is upgraded to its non-Hermitian analogue. Finally, we will break the SU(2) symmetry and discuss the anisotropic non-Hermitian Kondo effect. The results in this section are valid to two-loop order, and the details of the RG analysis are explained in Appendix.~\ref{app:rg_detail}. A summary containing a review of previous results as well as the new results we have obtained is in Table~\ref{tab:summary}.

\subsection{Isotropic case}
\subsubsection{Hermitian Kondo effect in Luttinger liquid}
\begin{figure}
    \centering
    \subfigure[]{
    \includegraphics[width=0.45\linewidth]{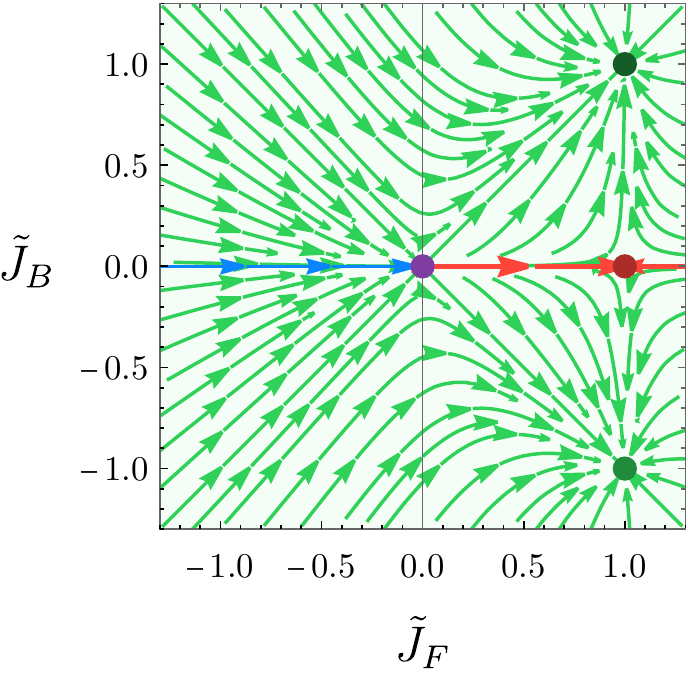}\label{fig:realLK0}}
    \subfigure[]{
    \includegraphics[width=0.45\linewidth]{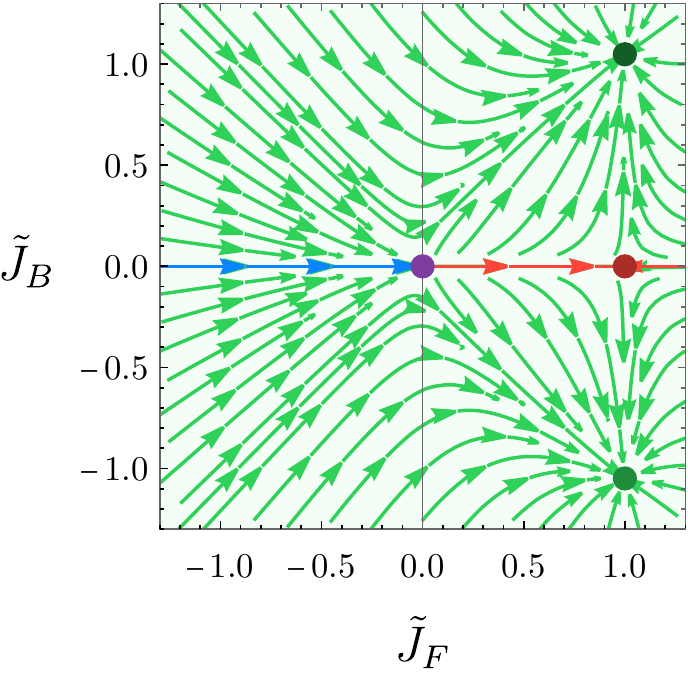}\label{fig:realLK1}}
    \caption{RG flow diagrams of the isotropic Hermitian Kondo effect in a Luttinger liquid for (a) $K_{c}=1$ and (b) $K_{c}=0.8$. 
    The purple and red dots stand for the (unstable) Gaussian fixed point and the two-channel fixed point which is unstable due to $J_{B}$. The green dots are stable, but their perturbative location in the figure should be replaced by the corresponding strong coupling limit. These dots exhibit non-Fermi liquid behavior when $K_{c}<1$, or Fermi liquid behavior when $K_{c}=1$.}
    \label{fig:realLK}
\end{figure}

Here, we review the isotropic SU(2) symmetric Hermitian Kondo effect in a Luttinger liquid, $J_{\perp F}=J_{z F}=J_{F}$ and $J_{\perp B}=J_{z B}=J_{B}$ in Eqs.~(\ref{eq:HFaniso}) and (\ref{eq:HBaniso}) \cite{furusaki1994,per1995,per1996}.
Before going further, we redefine the coupling constants as $\tilde{J}_{F/B}\equiv J_{F/B}/2\pi v_{F}$ and $\tilde{g}_{2}\equiv g_{2}/2\pi v_{F}$ for convenience.
The resulting RG flow equations in terms of $\tilde{J}_{F,B}$ and $\tilde{g}_{2}$ are given by \cite{furusaki1994}
\begin{align}
\frac{d\tilde{J}_{F}}{d\ell}={}&(\tilde{J}_{F}^{2}+\tilde{J}_{B}^{2})-\tilde{J}_{F}(\tilde{J}_{F}^{2}+\tilde{J}_{B}^{2}),\label{eq:JF}\\
\frac{d\tilde{J}_{B}}{d\ell}={}&\tilde{g}_{2}(1-\tilde{g}_{2})\tilde{J}_{B}+2\tilde{J}_{F}\tilde{J}_{B}-\tilde{J}_{B}(\tilde{J}_{F}^{2}+\tilde{J}_{B}^{2})\notag\\
\approx{}&\frac{1}{2}(1-K_{c})\tilde{J}_{B}+2\tilde{J}_{F}\tilde{J}_{B}-\tilde{J}_{B}(\tilde{J}_{F}^{2}+\tilde{J}_{B}^{2}).\label{eq:JB}
\end{align}
where we use the fact that $\frac{1}{2}(1-K_{c})\approx \tilde{g}_{2}(1-\tilde{g}_{2})$, $\ell=\ln(\Lambda/\mu)$, $\mu$ is the cutoff being continuously lowered, and $\Lambda$ is the original cutoff. Note that $\tilde{g}_{2}$ does not evolve under the RG flow, $\frac{d\tilde{g}_{2}}{d\ell}=0$, so we consider $K_{c}$ (or $\tilde{g}_{2}$) as a control parameter.
The RG flow diagrams for $K_{c}=1$ and $K_{c}=0.8$ are shown in Fig.~\ref{fig:realLK}.

The RG equations have three distinct fixed points. Two of these can be found on the line $\tilde{J}_{B}=0$, which is where we will focus first. 
On this line, the RG flow for an initially ferromagnetic coupling ($\tilde{J}_{F}<0$) goes to the Gaussian fixed point (blue line and purple dot in Fig.~\ref{fig:realLK}). However, in the antiferromagnetic case ($\tilde{J}_{F}>0$), the RG flow goes to a two-channel Kondo fixed point \cite{affleck1993} at $(\tilde{J}_{F}^{*},\tilde{J}_{B}^{*})=(1,0)$ (red line and red dot in Fig.~\ref{fig:realLK}). The two-channel fixed point is stable in dimensions $d > 1$ because backward Kondo scattering events are prohibited. 
In one dimension however, we see that the introduction of $\tilde{J}_{B}$ is important because $\tilde{J}_{B}$ is relevant for repulsive electron-electron interactions ($g_{2}>0$ or equivalently $K_{c}<1$) \cite{furusaki1994,per1995,per1996}.

In the general case where $\tilde{J}_{B}\neq0$, with ferromagnetic forward Kondo coupling ($\tilde{J}_{F}<0$), the RG flow monotonically decreases both $|\tilde{J}_{F}|$ and $|\tilde{J}_{B}|$. $\tilde{J}_{F}$ decreases to 0 earlier than $\tilde{J}_{B}$ and crosses over to the antiferromagnetic regime ($\tilde{J}_{F}>0$), hence $\tilde{J}_{B}$ cannot change its sign. Thus, even if starting with ferromagnetic forward Kondo coupling, it becomes antiferromagnetic under the RG flow.
Now that we are in the $\tilde{J}_{F}>0$ regime, the RG flow goes to another fixed point at $(\tilde{J}_{F}^{*},\tilde{J}_{B}^{*})=(1,\pm\sqrt{1+(1-K_{c})/2})$ (green dots in Fig.~\ref{fig:realLK}). It appears to be stable in the RG flow diagram, but it is known that the corresponding strong coupling limit ($J_{F}\rightarrow\infty,J_{B}\rightarrow\pm\infty$) is actually stable \cite{furusaki1994}. 
At this fixed point, $\tilde{J}_{B}$ can be either ferromagnetic ($\tilde{J}_{B}<0$) or antiferromagnetic ($\tilde{J}_{B}>0$).  
In the non-interacting limit $g_{2}=0$, this fixed point corresponds to the single-channel Kondo fixed point which gives us Fermi liquid behavior \cite{furusaki1994,per1995,per1996} because the two channels (left and right movers) can be combined into a single fermion at the fixed point \cite{per1996}. In the interacting case $g_{2}>0$, we can no longer combine the left and right movers and the fixed point shows non-Fermi liquid behavior, for example, the specific heat $C\propto T^{1/K_{c}-1}$ and conductance $G\propto T^{1/K_{c}-1}$ \cite{furusaki1994,per1995,per1996}. 
This behavior makes sense because the fixed point Hamiltonian can be interpreted as two semi-infinite Luttinger liquids and a spin singlet \cite{furusaki1994}.

\subsubsection{Non-Hermitian Kondo effect in Luttinger liquid}

\begin{figure}[t]
    \centering
    \subfigure[]{
    \includegraphics[width=0.45\linewidth]{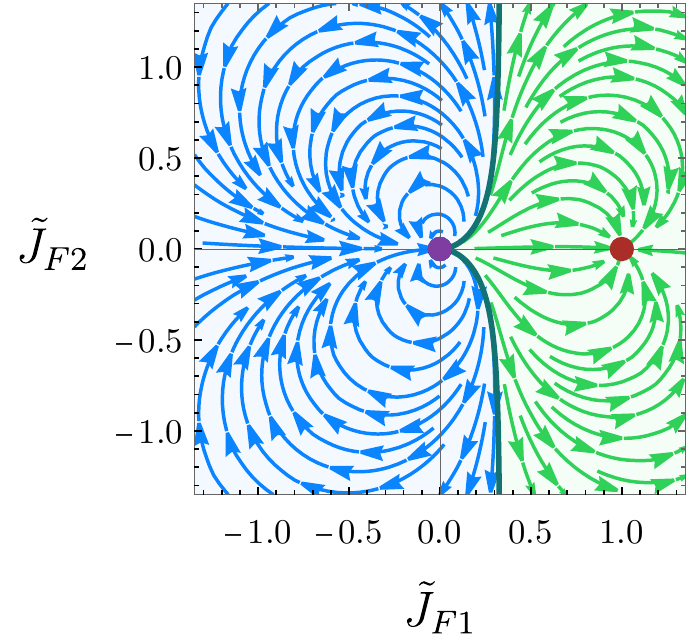}\label{fig:reversion}
    }
    \subfigure[]{
    \includegraphics[width=0.45\linewidth]{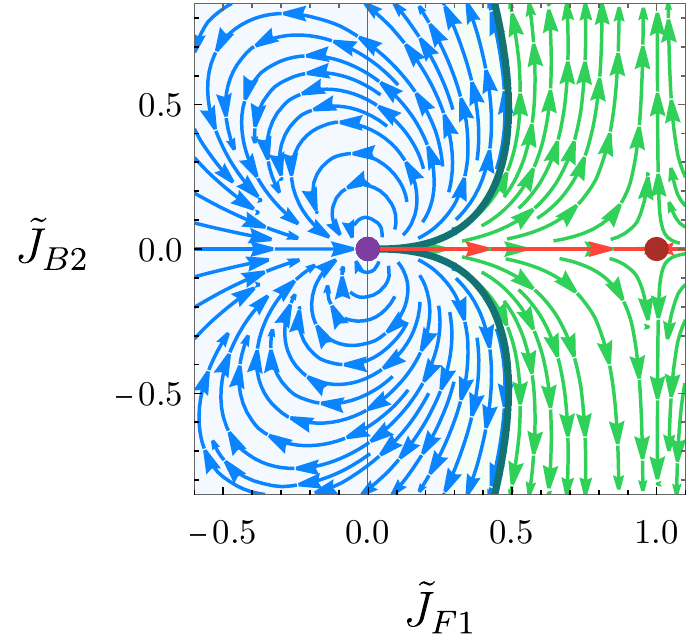}\label{fig:NHphase_a}
    }
    \subfigure[]{
    \includegraphics[width=0.45\linewidth]{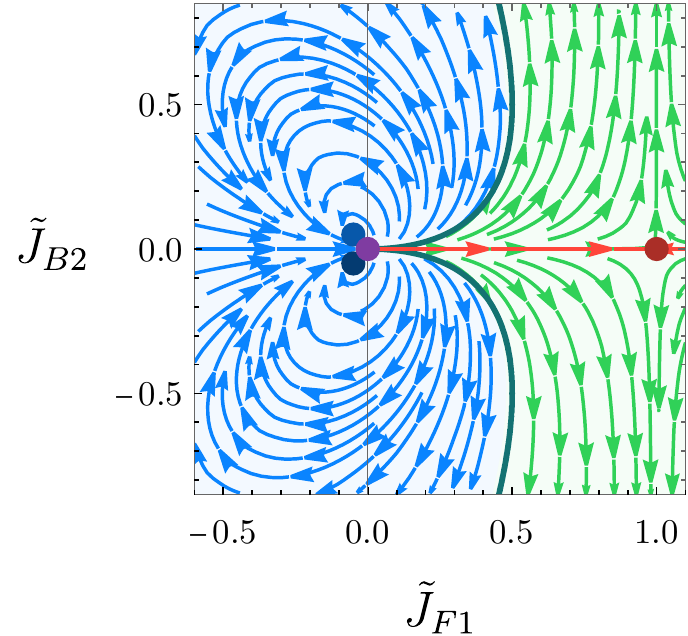}\label{fig:NHphase_c}
    }
    \subfigure[]{
    \includegraphics[width=0.45\linewidth]{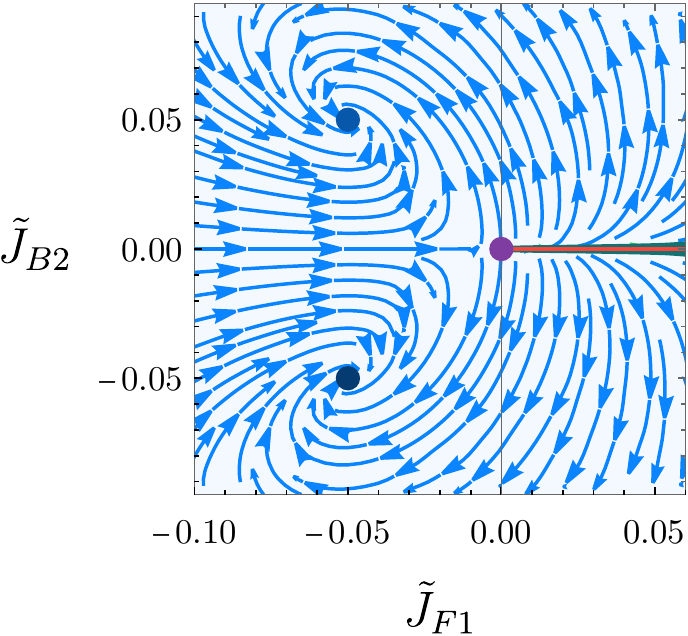}\label{fig:NHphase_d}
    }
    \caption{RG flow diagrams in non-Hermitian case. 
    (a) The RG flow diagram with $\tilde{J}_{B1}=\tilde{J}_{B2}=0$. The reversion of the RG flow appears. This diagram remains qualitatively the same  regardless of $\tilde{g}_2$ (including $\tilde{g}_2 = 0$). 
    (b) The RG flow for $K_{c}=1$ in terms of $\tilde{J}_{F1}$ and $\tilde{J}_{B2}$ with $\tilde{J}_{F2}=\tilde{J}_{B1}=0$. 
    (c-d)  The RG flow for $K_{c}=0.8$ in terms of $\tilde{J}_{F1}$ and $\tilde{J}_{B2}$ with $\tilde{J}_{F2}=\tilde{J}_{B1}=0$. The complex fixed points (blue dots) bifurcate from the Gaussian fixed point (purple dot) by decreasing $K_{c}$ from unity, and the RG flow shows spiral-like behaviors near the complex fixed points. The red dot stands for the two-channel Kondo fixed point.
    }
    \label{fig:NHphase}
\end{figure}

Here, we complexify the coupling constants of the previous section, keeping in mind that this complexification is due to dissipative processes from the Lindblad equation. This yields the isotropic non-Hermitian Hamiltonian with complex-valued Kondo couplings $\tilde{J}_{F}=\tilde{J}_{F1}+i\tilde{J}_{F2}$ and $\tilde{J}_{B}=\tilde{J}_{B1}+i\tilde{J}_{B2}$ where $\tilde{J}_{F1/F2}$ and $\tilde{J}_{B1/B2}$ are real. The RG flow equations for $\tilde{J}_{F1},\tilde{J}_{F2},\tilde{J}_{B1},\tilde{J}_{B2}$ are given by
\begin{align}
\frac{d\tilde{J}_{F1}}{d\ell}={}&(1-\tilde{J}_{F1})(\tilde{J}_{F1}^{2}-\tilde{J}_{F2}^{2}+\tilde{J}_{B1}^{2}-\tilde{J}_{B2}^{2})\notag\\
&+2\tilde{J}_{F2}(\tilde{J}_{F1}\tilde{J}_{F2}+\tilde{J}_{B1}\tilde{J}_{B2}),\label{eq:JF1}\\
\frac{d\tilde{J}_{F2}}{d\ell}={}&2(1-\tilde{J}_{F1})(\tilde{J}_{F1}\tilde{J}_{F2}+\tilde{J}_{B1}\tilde{J}_{B2})\notag\\
&-\tilde{J}_{F2}(\tilde{J}_{F1}^{2}-\tilde{J}_{F2}^{2}+\tilde{J}_{B1}^{2}-\tilde{J}_{B2}^{2}),\\
\frac{d\tilde{J}_{B1}}{d\ell}={}&\frac{1}{2}(1-K_{c})\tilde{J}_{B1}+2(\tilde{J}_{F1}\tilde{J}_{B1}-\tilde{J}_{F2}\tilde{J}_{B2})\notag\\&-\tilde{J}_{B1}(\tilde{J}_{F1}^{2}-\tilde{J}_{F2}^{2}+\tilde{J}_{B1}^{2}-\tilde{J}_{B2}^{2})\notag\\&+2\tilde{J}_{B2}(\tilde{J}_{F1}\tilde{J}_{F2}+\tilde{J}_{B1}\tilde{J}_{B2}),\\
\frac{d\tilde{J}_{B2}}{d\ell}={}&\frac{1}{2}(1-K_{c})\tilde{J}_{B2}+2(\tilde{J}_{F1}\tilde{J}_{B2}+\tilde{J}_{F2}\tilde{J}_{B1})\notag\\&-2\tilde{J}_{B1}(\tilde{J}_{F1}\tilde{J}_{F2}+\tilde{J}_{B1}\tilde{J}_{B2})\notag\\&-\tilde{J}_{B2}(\tilde{J}_{F1}^{2}-\tilde{J}_{F2}^{2}+\tilde{J}_{B1}^{2}-\tilde{J}_{B2}^{2}).\label{eq:JB2}
\end{align}
The RG flow diagrams are shown in Fig.~\ref{fig:NHphase}.

In the RG equations, when we set $\tilde{J}_{B1}=\tilde{J}_{B2}=0$, regardless of the value of $\tilde{g}_{2}$, we observe a reversion of the RG flow similar to the non-Hermitian Kondo effect in higher dimensions \cite{Nakagawa2018} (Fig.~\ref{fig:reversion}). 
Thus, even if the initial Kondo coupling is antiferromagnetic ($\tilde{J}_{F1}>0$), it is possible that the RG flow approaches the non-Kondo phase which is represented by the Gaussian fixed point; this type of flow violates the $g$ theorem \cite{Nakagawa2018,Lourenco2018}. However, when we turn on the backward scatterings $\tilde{J}_{B1/B2}$, the situation changes. In the non-interacting limit ($g_{2}=0$), we still see the reversion of the RG flow (Fig.~\ref{fig:NHphase_a}), and the RG flow in the non-Kondo phase (blue region in Fig.~\ref{fig:NHphase_a}) goes to the Gaussian fixed point. However, in the interacting ($g_{2}>0$) case, we find a pair of stable fixed points in the perturbative regime, whose fixed point values are $(\tilde{J}_{F1}^{*},\tilde{J}_{F2}^{*},\tilde{J}_{B1}^{*},\tilde{J}_{B2}^{*})=(-\frac{1}{4}(1-K_{c}),0,0,\pm\frac{1}{4}(1-K_{c}))$. For simplicity, we illustrate the flow to this fixed point for the case $\tilde{J}_{F2}=\tilde{J}_{B1}=0$. These new (complex) points bifurcate from the Gaussian fixed point into the complex parameter plane as we decrease $K_{c}$ from unity (see Figs.~\ref{fig:NHphase_c} and \ref{fig:NHphase_d}). 

These complex fixed points are not analogous to any multichannel problem because the backward scattering is nonzero, and the forward scattering is ferromagnetic. This is unusual because typically Kondo fixed points are typically antiferromagnetic. Furthermore, they are in the perturbative regime, and therefore a strong coupling analysis is unnecessary. The fact that the fixed point values depend on $K_{c}$ (or $\tilde{g}_{2}$) conveys that the complex fixed points arise from the interplay between non-Hermitian Kondo interactions and electron-electron interaction.
In addition, the complex fixed points have complex scaling dimensions, $\Delta\approx
\frac{e^{\pm i\pi/3}}{2}(1-K_{c})$. 
The spiral-like RG flow near the complex fixed point is observed due to the imaginary part of the scaling dimension; accordingly, the complex fixed point is of the stable-focus type \cite{Calabrese2002,Calabrese2003,KHMELNITSKII197859,Dorogovtsev1980,Yerzhakov2018,Yerzhakov2021}.
The real part of the scaling dimension determines how fast the RG flow approaches the fixed point, while the imaginary part determines the period of the cycle for the spiral-like behaviors.
The stabilities of other fixed points from the Hermitian case remain unchanged, so the Gaussian fixed point and two-channel Kondo fixed point are still unstable but the isotropic (real) strong coupling limit fixed point is stable as before. 
Note that the results make sense within the context of fewer loops as well, because often a divergence of a coupling at lower orders in perturbation theory will produce a finite fixed point at higher orders. 
For example, at tree level, $\tilde{J}_{B1/B2}$ diverges in the presence of non-zero $\tilde{g}_{2}$, but $\tilde{J}_{F1/F2}$ are marginal. 
At one-loop order, we find the complex fixed points arising from the competition between the relevant bare scaling dimension of the imaginary backward Kondo coupling $\tilde{J}_{B2}$ (depending on the electron-electron interaction $\tilde{g}_{2}$), and the one-loop contribution. However, one-loop order is not enough to find the Hermitian Kondo fixed points; for example, the two-channel Kondo fixed point does not exist because the forward Kondo couplings diverge at this order.
At two-loop order, the complex fixed points survive and have the same fixed point values and scaling dimension in terms of $K_{c}$, but we also get the (Hermitian) two-channel Kondo fixed point. Thus, two-loop order is required to observe both the complex fixed points and Hermitian Kondo fixed points.
Another remark is that, due to the presence of the imaginary Kondo couplings, our model explicitly breaks $\mathcal{PT}$ symmetry. The imaginary Kondo interactions are even under parity, but odd under time reversal. This can be compared with the $\mathcal{PT}$-symmetric case without electron-electron interactions, in which one finds spiral behavior of the RG flow only when the $\mathcal{PT}$ symmetry is spontaneously broken and the energy eigenvalues subsequently become complex \cite{Lourenco2018}. The presence of a qualitatively similar spiral RG flow can likely be attributed to the breaking of $\mathcal{PT}$ symmetry, whether it occurs spontaneously or explicitly.

\begin{figure*}
    \centering
    \subfigure[]{\includegraphics[width=0.31\linewidth]{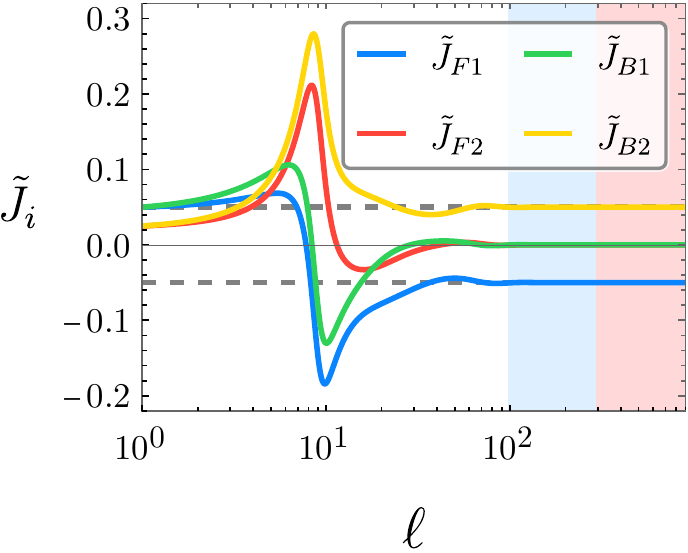}\label{fig:isoRGflow}
    }  
    \subfigure[]{\includegraphics[width=0.31\linewidth]{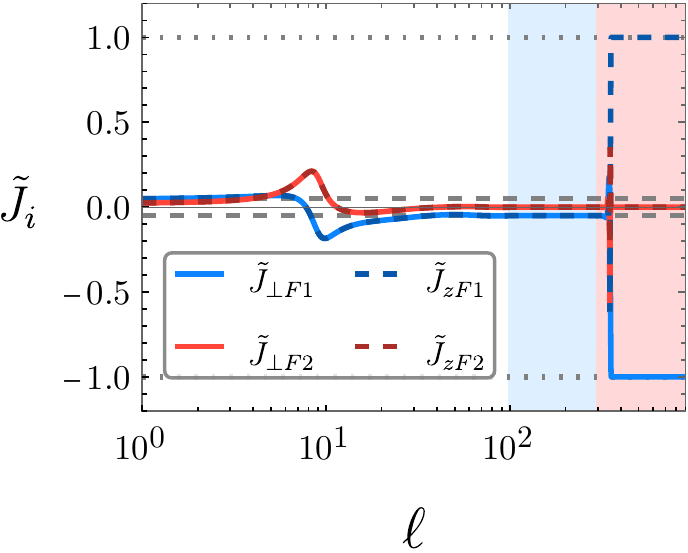}\label{fig:anisotropic1}
    }  
    \subfigure[]{\includegraphics[width=0.31\linewidth]{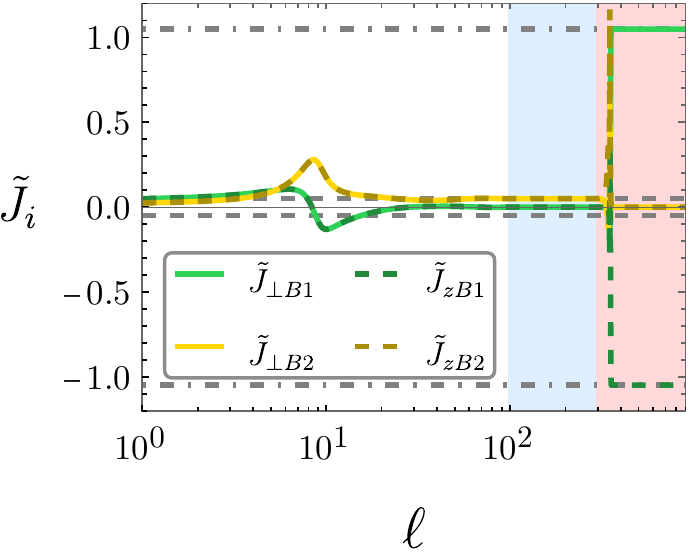}\label{fig:anisotropic2}
    }
    \caption{RG flow of the isotropic and anisotropic non-Hermitian cases for initial values $\tilde{J}_{\perp F1}=\tilde{J}_{z F1}=\tilde{J}_{\perp B1}=\tilde{J}_{\perp B1}=0.05$ and $\tilde{J}_{\perp F2}=\tilde{J}_{z F2}=\tilde{J}_{\perp B2}=\tilde{J}_{\perp B2}=0.025$ with $K_{c}=0.8$. (a) The RG flow under the isotropic RG equations goes to the complex fixed point, $(\tilde{J}_{F1},\tilde{J}_{F2},\tilde{J}_{B1},\tilde{J}_{B2})=(-0.05,0,0,0.05)$. (b-c) The RG flow under the anisotropic RG equations.
    In the blue-shaded region ($100\lesssim\ell\lesssim300$), the RG flow stays near the complex fixed point but in the red-shaded region ($300\lesssim\ell$), it eventually goes to the isotropic strong coupling limit, $-\tilde{J}_{\perp F1}^{*}=\tilde{J}_{z F1}^{*}=1$, $-\tilde{J}_{\perp B1}^{*}=\tilde{J}_{z B1}=1.0488$, and $\tilde{J}_{\perp F2}^{*}=\tilde{J}_{z F2}^{*}=\tilde{J}_{\perp B2}^{*}=\tilde{J}_{z B2}^{*}=0$.
    The gray dashed, dotted, and dot-dashed lines stand for values of $\pm0.05$, $\pm1$, and $\pm1.0488$, respectively.
    }
    \label{fig:anisoiso}
\end{figure*}

\subsubsection{Critical properties of Complex fixed point}
The scaling dimension of the local impurity correlation function  determines the critical properties of the complex fixed points. 
In non-Hermitian quantum theory, the right and left eigenstates $\bra{n_{L}}$ and $\ket{n_{R}}$ which satisfy $H_{\text{NH}}\ket{n_{R}}=E_{n}\ket{n_{R}}$ and $H_{\text{NH}}^{\dagger}\ket{n_{L}}=E_{n}^{*}\ket{n_{L}}$ are not orthogonal but biorthogonal, $\braket{n_{L}|m_{R}}\propto\delta_{nm}$ and $\braket{n_{L}|m_{L}},\braket{n_{R}|m_{R}}\neq\delta_{nm}$ \cite{Brody2014}. 
Thus, we can compute two types of correlation functions: the biorthogonal correlation function $\braket{\mathcal{O}}_{LR}\equiv \braket{0_{L}|\mathcal{O}|0_{R}}/\braket{0_{L}|0_{R}}$ obtained by path integrals \cite{Yamamoto2019,Yamamoto2022}, and the right-state correlation function $\braket{\mathcal{O}}_{RR}\equiv \braket{0_{R}|\mathcal{O}|0_{R}}/\braket{0_{R}|0_{R}}$ obtained by the wave-functional approach \cite{Ashida2016,Yamamoto2022,Furukawa2011}. Here, $\bra{0_{L}}$ and $\ket{0_{R}}$ are the left and right ground states, where the real part of the eigenenergy is the lowest.
Since our RG analysis is based on the path integral approach, the impurity correlation function obtained here is the biorthogonal correlation function $\chi_{LR}(\tau)=\braket{S(\tau)S(0)}_{LR}$. 
The correlation function near the complex fixed points will behave as $\chi_{LR}(\tau)\propto |\tau|^{-2\Delta_{LR}}$ with $\Delta_{LR}=\frac{e^{\pm i\pi/3}}{2}(1-K_{c})$. Since the scaling dimension of the complex fixed point is complex-valued, the correlation function will simultaneously show two types of behavior: oscillatory behavior as a function of $\log\tau$, whose periodicity is determined by the imaginary part of the scaling dimension, and also power-law scaling which is determined by the real part of the scaling dimension \cite{KHMELNITSKII197859,Dorogovtsev1980,Aharony2018,Yerzhakov2018,Gorbenko2018b,Yerzhakov2021}.
Note that since we are working in an open system, the inverse temperature in the partition function $\mathcal{Z}\propto \sum_{n}e^{-\beta E_{n}}=\sum_{n}\braket{n_{L}|e^{-\beta H_{\text{eff}}}|n_{R}}$  can only be interpreted as a parameter, not a physical temperature. Thus we consider $\tau$ simply as a parameter to define scaling behaviors of the physical quantities, and the limit of infinite $\beta$ to define the ground states (defined as those states whose eigenvalues have the lowest real part) \cite{Ashida2016,Yamamoto2019,Yamamoto2022}.

\begin{table*}[t]
    \centering
    \begin{tabular}{>{$}c<{$}||c|c||c|c}
        \text{Cases} & FPs in Hermitian Kondo & Properties & FPs in non-Hermitian Kondo & Properties \\ \hline\hline
        \begin{tabular}{>{$}c<{$}}J_{B}=0\\\text{(isotropic)}\end{tabular} & \begin{tabular}{c}
             Gaussian \\
             2-Channel Kondo
        \end{tabular}&\begin{tabular}{c}
             Unstable \\
             Stable
        \end{tabular}&\begin{tabular}{c}
             Gaussian \\
             2-Channel Kondo
        \end{tabular}&\begin{tabular}{c}
             Unstable, reversion of RG flow \\
             Stable
        \end{tabular}\\ \hline
        \begin{tabular}{>{$}c<{$}}g_{2}=0,\\J_{B}\neq0\\\text{(isotropic)}\end{tabular} & \begin{tabular}{c}
             Gaussian \\
             2-Channel Kondo\\
             1-Channel Kondo
        \end{tabular}&\begin{tabular}{c}
             Unstable \\
             Unstable \\
             Stable SCL (FL) 
        \end{tabular}&\begin{tabular}{c}
             Gaussian \\
             2-Channel Kondo\\
             1-Channel Kondo
        \end{tabular}&\begin{tabular}{c}
             Unstable, reversion of RG flow \\
             Unstable\\
             Stable SCL (FL)
        \end{tabular}\\ \hline
        \begin{tabular}{>{$}c<{$}}g_{2}\neq0,\\J_{B}\neq0\\\text{(isotropic)}\end{tabular} & \begin{tabular}{c}
             Gaussian \\
             2-Channel Kondo\\
             Isotropic strong coupling\\{}
        \end{tabular}&\begin{tabular}{c}
             Unstable \\
             Unstable \\
             Stable SCL (NFL) \\{}
        \end{tabular}&\begin{tabular}{c}
             Gaussian \\
             2-Channel Kondo\\
             Isotropic strong coupling\\
             \emph{Complex fixed point}
        \end{tabular}&\begin{tabular}{c}
             Unstable\\
             Unstable\\
             Stable SCL (NFL)\\
             Stable; Spiral-like RG flow
        \end{tabular}\\ \hline
        \begin{tabular}{>{$}c<{$}}g_{2}\neq0,\\J_{B}\neq0\\\text{(anisotropic)}\end{tabular} & \begin{tabular}{c}
             Gaussian \\
             2-Channel Kondo\\
             Ising fixed line\\
             Isotropic strong coupling\\{}
        \end{tabular}&\begin{tabular}{c}
             Unstable \\
             Unstable \\
             Stable\\
             Stable SCL (NFL) \\{}
        \end{tabular}&\begin{tabular}{c}
             Gaussian \\
             2-Channel Kondo\\
             \emph{Ising fixed plane} \\
             Isotropic strong coupling\\
             \emph{Complex fixed point}
        \end{tabular}&\begin{tabular}{c}
             Unstable\\
             Unstable\\
             Stable\\
             Stable SCL (NFL) \\
             Unstable; Walking behavior
        \end{tabular}\\
    \end{tabular}
    \caption{A table of both the fixed points found in previous work, as well as our new results, to easier contrast the two. For the non-Hermitian case, ``$J_{B}\neq0$'' means that the real backward scattering is nonzero for the first column, but both real and imaginary backward scatterings are nonzero for the second column. The fixed points in the Hermitian Kondo and non-Hermitian Kondo cases, when they have the same name, are the same fixed point, due to the imaginary part (in the non-Hermitian case) flowing to zero and leading to a Hermitian fixed point Hamiltonian. The fixed points in italics have non-zero imaginary values for the couplings. In particular, the difference between the Ising fixed line and \emph{Ising fixed plane} (with strong backward-scattering Kondo coupling) is that the latter becomes a plane due to the imaginary component of $J_{zF2}$ coupling. We emphasize that both the non-Hermitian backward scattering and electron-electron interactions are necessary to produce a complex fixed point. The following abbreviations FPs = fixed points, SCL = strong coupling limit, FL = Fermi liquid, NFL = non-Fermi liquid.}
    \label{tab:summary}
\end{table*}

\subsection{Anisotropic case}
We now introduce anisotropy into the system by allowing the Kondo couplings for the $x$ and $y$ spin components to differ from that of the $z$ component. Let us review the anisotropic Hermitian case first, the RG flow equations for which are given by Eqs.~(\ref{eq:JpFn2})-(\ref{eq:g2n2}). Introducing the anisotropy has different effects depending on whether or not there is backward scattering. If there is no backward scattering, there is a new (stable) Ising fixed line, where $J_{\perp F}^{*} = 0$ and $J_{z F}^{*}<0$. The RG flow either goes towards this line or towards the isotropic two-channel fixed point depending on whether $J_{z F}<-|J_{\perp F}|$ or $J_{z F}>-|J_{\perp F}|$, respectively. 
However, if there is also the backward Kondo scattering, then introducing the anisotropy destabilizes the Ising fixed line along $J_{\perp B}=J_{z B}=0$ but creates a different stable Ising fixed line. This time it is at strong coupling with infinite backward Ising coupling, at $J_{z F}^{*}<0$ and $J_{z B}^{*}\rightarrow\infty$ with $J_{\perp F}^{*}=J_{\perp B}^{*}=0$. Only in the region $J_{z F}<0$ may the flow approach this strong coupling fixed line, and all other flows will tend towards the isotropic strong coupling fixed point, $(\tilde{J}_{\perp F}^{*},\tilde{J}_{z F}^{*},\tilde{J}_{\perp B}^{*},\tilde{J}_{z B}^{*})=(ab,1,a\sqrt{1+\tfrac{1}{2}(1-K_{c})},b\sqrt{1+\tfrac{1}{2}(1-K_{c})})$ with $a,b=\pm1$.
In all of these cases, the anisotropy is irrelevant when $J_{z F} > 0$, but (at least initially) relevant when $J_{z F} < 0$ \cite{Anderson1970,Hewson1993}, although the flow may eventually reach an isotropic point. 

In the non-Hermitian case, the spirit of the anisotropy remains the same: anisotropy is irrelevant when $J_{z F1} > 0$, but (at least initially) relevant when $J_{z F1} < 0$. 
Similar to the anisotropic Hermitian case, the anisotropy introduces a stable Ising fixed plane with strong backward-scattering Kondo coupling, at $J_{z F1}^{*}<0$, $J_{z F2}^{*}\in\mathbb{R}$ and $|J_{z B1}|,|J_{z B2}|\rightarrow\infty$ with $\tilde{J}_{\perp F1}=\tilde{J}_{\perp F2}=\tilde{J}_{\perp B1}=\tilde{J}_{\perp B2}=0$; this fixed plane however also contains strong \textit{imaginary-valued} backward Kondo couplings in contrast to the Hermitian case.
The more interesting case is the fate of the complex fixed point under anisotropy. Due to the ferromagnetic nature of the complex fixed point, this suggests that the anisotropy will destabilize it. This is indeed the case, and it is destabilized. Now, the RG flow approaches the complex fixed point and exhibits walking behavior in its vicinity \cite{Piai2010,Gorbenko2018a,Gorbenko2018b,ma2019,ma2020}, but eventually flows to the isotropic strong coupling limit or Ising fixed plane with strong backward-scattering Kondo coupling. 
To exemplify the walking behavior around the (now unstable) complex fixed point, for the initial values $\tilde{J}_{\perp F1}=\tilde{J}_{z F1}=\tilde{J}_{\perp B1}=\tilde{J}_{\perp B1}=0.05$ and $\tilde{J}_{\perp F2}=\tilde{J}_{z F2}=\tilde{J}_{\perp B2}=\tilde{J}_{\perp B2}=0.025$ with $K_{c}=0.8$, the RG flow under the isotropic RG flow equations goes to the complex fixed point (Fig.~\ref{fig:isoRGflow}). The RG flow under the anisotropic RG flow equations goes the complex fixed point around about $\ell\approx100$ and stays between $100\lesssim\ell\lesssim300$ (blue-shaded region in Fig.~\ref{fig:anisotropic1} and \ref{fig:anisotropic2}), but after about $\ell\approx300$, the RG flow goes to the isotropic strong coupling limit fixed point (red-shaded region in Fig.~\ref{fig:anisotropic1} and \ref{fig:anisotropic2}).

A summary of all the results in isotropic and anisotropic cases is listed in Table~\ref{tab:summary}.

\section{Discussion}\label{sec:conclusion}
In this work, we studied the non-Hermitian Kondo effect in a Luttinger liquid.
By considering non-Hermitian forward and backward Kondo interactions with weak repulsive electron-electron interactions, we find a pair of novel complex fixed points with complex scaling dimensions which are stable in the isotropic limit. These fixed points arise from the interplay between the electron-electron interactions and non-Hermitian Kondo interactions. 
In contrast to usual fixed points with the real-valued scaling dimension, near the complex fixed points, due to the complex-valued scaling dimension, the RG flow shows spiral-like behaviors, in addition, the biorthogonal impurity correlation function would exhibit oscillatory behaviors as well as power-law behaviors.
In the generic anisotropic case, the RG flow approaches very close to the complex fixed point and stays for a while (walking behavior), but eventually goes to the isotropic strong coupling limit or Ising fixed plane with strong backward-scattering Kondo coupling.

Such complex fixed points and surrounding spiral-like RG flow are also observed in the complexified Potts model with $N>4$ and deconfined pseudocriticality \cite{Gorbenko2018a,Gorbenko2018b,ma2019,ma2020}.
The complex fixed point in such cases is introduced to explain the origin of the weakly first-order phase transition in the corresponding Hermitian model, and the transition is characterized by an extremely long correlation length compared to the lattice constant.
However, in our work, the non-Hermitian interaction is explicitly introduced by the Lindblad equation due to interaction with the environment. Accordingly, the complex fixed point in our study may actually be accessible experimentally.

The properties of this complex fixed point can be experimentally investigated in the realm of cold atoms. By imposing spin-space isotropy, this provides a platform to observe a system right at a complex fixed point. Upon breaking the isotropy, it would be possible to observe walking behavior, as the system approaches and then leaves the complex fixed point, as the energy scale of the system is lowered. This allows measurement of the critical properties of the complex fixed point.

Future work for this model would mainly consist of an exact solution of the scaling dimension at the complex field point. In the case of the Hermitian Kondo model, its exact solution can be obtained through boundary conformal field theory (CFT). 
A similar CFT solution also exists for the Hermitian Kondo effect in a Luttinger liquid, in which case the scaling dimension of the leading irrelevant operator is $1 + \Delta = \frac{1}{2}(K_{c}^{-1}+1)$ \cite{furusaki1994,per1995,per1996}.
Thus, to obtain the exact scaling dimension, we will need to apply boundary CFT analysis \cite{affleck1993,Ludwig1994,per1995,per1996,Patri2020} or the Bethe-ansatz approach \cite{Andrei1980,Wiegmann1981,Andrei1984,Andrei1983,Tsvelick1985,Tsvelick1984,Nakagawa2018} for the complex fixed point.

Another direction for future work would be numerical analysis. As mentioned before, in non-Hermitian quantum theory, both biorthogonal correlation functions and right-state correlation functions are important.
The scaling behaviors of the biorthogonal and right-state correlation functions are generically different \cite{Ashida2016,Yamamoto2022,Yamamoto2023} but there is no known direct way to obtain one from the other.
In our case, we were able to extract the scaling behavior of the biorthogonal correlation function, not the right-state correlation function.
To study the scaling behavior of the right-state correlation function, numerical computation will be needed, such as exact diagonalization or density matrix renormalization group methods \cite{Yamamoto2022}. By computing and comparing the dynamic impurity correlation functions from the biorthogonal and right-state correlation functions, we may find a relation between them. Furthermore, since the right-state correlation function is more closely related to the physical observables \cite{Yamamoto2022}, its structure would be helpful in understanding the critical properties of the complex fixed point. 

In conclusion, our work reveals that complex fixed points can arise in open quantum systems through electron-electron interactions, offering a new example of the interplay between non-Hermitian physics and electron-electron interactions. These findings open the door to exciting possibilities for making connections between complex conformal field theory and cold-atom experiments.

\begin{acknowledgements}
This work was supported by the NSERC Of Canada and Center for Quantum Materials at the University of Toronto. Y.B.K.~is further supported by the Simons Fellowship from the Simons Foundation and the Guggenhein Fellowship from the John Simon Guggenheim Memorial Foundation. Y.B.K.~acknowledges the support of the Advanced Study Group on ``Entanglement and Dynamics of Quantum Matter" at the Center for Theoretical Physics of Complex Systems in the Institute for Basic Science, where a part of the current work was done. D.J.S.~is supported by the Ontario Graduate Scholarship.
\end{acknowledgements}

\appendix
\section{Non-Hermitian Hamiltonian from Lindblad equation}\label{app:NH_Kondo}
From the Lindblad equation, the non-Hermitian part of the effective Hamiltonian is given by
\begin{align}
H_{\text{AH}}=-\frac{i}{2}\sum_{i}L_{i}^{\dagger}L_{i},
\end{align}
where $L_{i}$'s are jump operators, and $i = +,-,\uparrow,\downarrow$. In this work, we consider the jump operators $L_{i}$ for the two-body losses \cite{Nakagawa2018},
\begin{align}
L_{\pm}={}&\frac{\gamma_{\pm}}{\sqrt{2}}( \Psi_{\uparrow,0}f_{\downarrow}\pm\Psi_{\downarrow,0}f_{\uparrow}),\\
L_{\uparrow}={}&\gamma_{0} \Psi_{\uparrow,0}f_{\uparrow},\\
L_{\downarrow}={}&\gamma_{0} \Psi_{\downarrow,0}f_{\downarrow},
\end{align}
where $\Psi_{0}$ is a two-component conduction electron spinor located at the origin, and $f$ is a pseudofermion representation of the localized impurity (also located at the origin) which is defined by $\vec{S}=\sum_{\alpha,\beta=\uparrow,\downarrow}(f^{\dagger}_{\alpha}\frac{\vec{\sigma}_{\alpha\beta}}{2}f_{\beta})$ with $\sum_{\alpha=\uparrow,\downarrow}f_{\alpha}^{\dagger}f_{\alpha}=1$. 
Then, the $H_{\text{AH}}$ is
\begin{align}
H_{\text{AH}}=-\frac{i}{2}\sum_{j}L_{j}^{\dagger}L_{j}=i\sum_{j}v_{j}(\Psi_{0}^{\dagger}\frac{\tau^{j}}{2}\Psi_{0})(f^{\dagger}\frac{\sigma^{j}}{2}f),
\end{align}
where $v_{x}=v_{y}=\frac{1}{2}(\gamma_{-}^{2}-\gamma_{+}^{2})$, $v_{z}=\frac{1}{2}(\gamma_{-}^{2}+\gamma_{+}^{2}-2\gamma_{0}^{2})$, and $v_{0}=-\frac{1}{2}(\gamma_{-}^{2}+\gamma_{+}^{2}+2\gamma_{0}^{2})$. By using $\Psi_{0}\approx \psi_{L,0}+\psi_{R,0}$ and the definition of the pseudofermion, we can get the imaginary forward and backward scattering Kondo interactions.
Note that this also generates a potential scattering $v_{0}$, but we ignore it in the main text 
because it does not qualitatively change the results.

\section{Details of RG analysis}\label{app:rg_detail}

In this section, we give the details of our RG analysis. We use the dimensional regularization with the minimal subtraction scheme \cite{zhu2002}. For the field-theoretical calculation, we introduce the Abrikosov pseudofermion for the impurity spin, $\vec{S}=\sum_{\alpha,\beta=\uparrow,\downarrow}(f^{\dagger}_{\alpha}\frac{\vec{\sigma}_{\alpha\beta}}{2}f_{\beta})$, and the imaginary chemical potential $\lambda\sum_{\alpha=\uparrow,\downarrow}f^{\dagger}_{\alpha}f_{\alpha}$ to satisfy the physical constraint $\sum_{\alpha=\uparrow,\downarrow}f^{\dagger}_{\alpha}f_{\alpha}=1$. At the end of the calculation, we will take the limit, $\lambda\rightarrow\infty$ to recover the physical impurity Hilbert space.

The propagators are given by
\begin{align}
G_{(L,R),ij}(\omega,E_{p})={}&\frac{\delta_{ij}}{\omega\pm E_{p}},\\
G_{f,ij}(\omega)={}&\frac{\delta_{ij}}{\omega-\lambda},
\end{align}
and the density of state of the left- and right-moving fermions is given by $\sum_{p}\delta(\omega-E_{p})=N_{0}|\omega|^{-\epsilon'}$ with $N_{0}=1/(2\pi v_{F})$, where we introduce $\epsilon'$ to perform the dimensional regularization. At the end of the calculation, we will take $\epsilon'\rightarrow0$. 
The bare coupling constants are
\begin{align}
J_{i}^{\text{B}}={}&J_{i}Z_{f}^{-1}Z_{c}^{-1}Z_{J_{i}}\mu^{\epsilon'},\\
g_{2}^{\text{B}}={}&g_{2}Z_{c}^{-2}Z_{g_{2}}\mu^{\epsilon'}
\end{align}
where the superscript $\text{B}$ stands for the bare values, $Z_{g_{2}}$ and $Z_{J_{i}}$ are the renormalization constants for $g_{2}$ and $J_{i}$ ($i={\perp} F,zF,{\perp} B,zB$), respectively, and $Z_{c}$ and $Z_{f}$ are the renormalization constants for the left/right moving fermion and pseudofermion fields, respectively.

By computing the loop corrections up to two-loop order, we can get the renormalization constants. More details including the Feynman diagrams are presented in Ref.~\cite{zhu2002,han2022}. The renormalization constants are given by
\begin{widetext}
\begin{align}
(Z_{\perp F}-1)={}&-\frac{N_{0}}{\epsilon'}(J_{z F}+J_{\perp B}J_{z B}/J_{\perp F})
+\frac{N_{0}^{2}}{8\epsilon'}(J_{z F}^{2}+J_{z B}^{2}-4g_{2}^{2})\notag\\
&-\frac{2N_{0}^{2}}{\epsilon'^{2}}(J_{\perp F}^{2}+J_{z F}^{2}+J_{\perp B}^{2}+J_{z B}^{2})-\frac{2(N_{0}^{2}J_{\perp B}/J_{\perp F})}{\epsilon'^{2}}(2J_{\perp F}J_{\perp B}+2J_{z F}J_{z B}+g_{2}J_{z B}),\\
(Z_{z F}-1)={}&-\frac{N_{0}}{\epsilon'}(J_{\perp F}^{2}+J_{\perp B}^{2})/J_{z F}
+\frac{N_{0}^{2}}{8\epsilon'}(2J_{\perp F}^{2}+2J_{\perp B}^{2}-J_{z F}^{2}-J_{z B}^{2}-4g_{2}^{2})\notag\\
&-\frac{3N_{0}^{2}}{\epsilon'^{2}}(J_{\perp B}^{2}+J_{\perp F}^{2})-\frac{2N_{0}^{2}(J_{\perp B}/J_{z F})}{\epsilon'^{2}}(3J_{\perp F}J_{z B}+g_{2}J_{\perp B}),\\
(Z_{\perp B}-1)={}&-\frac{N_{0}}{\epsilon'}(g_{2}+J_{z F}+J_{\perp F}J_{z B}/J_{\perp B})
+\frac{N_{0}^{2}}{8\epsilon'}(J_{z F}^{2}+J_{z B}^{2})
-\frac{N_{0}^{2}}{\epsilon'^{2}}[(J_{\perp F}+J_{\perp B})^{2}+(J_{z F}+J_{z B})^{2}+g_{2}(g_{2}+2J_{z F})]\notag\\
&-\frac{(N_{0}^{2}J_{\perp F}/J_{\perp B})}{\epsilon'^{2}}[(J_{\perp F}+J_{\perp B})^{2}+(J_{z F}+J_{z B})^{2}+2g_{2}J_{z B}]
,\\
(Z_{z B}-1)={}&-\frac{N_{0}}{\epsilon'}(g_{2}J_{z B}+2J_{\perp F}J_{\perp B})/J_{z B}
+\frac{N_{0}^{2}}{8\epsilon'}(2J_{\perp F}^{2}+2J_{\perp B}^{2}-J_{z F}^{2}-J_{z B}^{2})\notag\\
&-\frac{N_{0}^{2}}{\epsilon'^{2}}[g_{2}^{2}+2(J_{\perp F}^{2}+J_{\perp F}J_{\perp B}+J_{\perp B}^{2})]-\frac{N_{0}^{2}}{\epsilon'^{2}}[(J_{z F}/J_{z B})(J_{\perp F}^{2}+4J_{\perp F}J_{\perp B}+J_{\perp B}^{2})+4g_{2}J_{\perp F}J_{\perp B}/J_{z B}],\\
(Z_{g_{2}}-1)={}&-\frac{N_{0}^{2}g_{2}^{2}}{\epsilon'},\\
(Z_{f}-1)={}&-\frac{N_{0}^{2}}{8\epsilon'}(2J_{\perp F}^{2}+2J_{\perp B}^{2}+J_{z F}^{2}+J_{z B}^{2}),\\
(Z_{c}-1)={}&-\frac{N_{0}^{2}g_{2}^{2}}{2\epsilon'}.
\end{align}
\end{widetext}
The RG flow equations are obtained by using the renormalization constants,
\begin{align}
\frac{dJ_{i}}{d\ell}={}&-J_{i}\bigg[\sum_{j}J_{j}\partial_{J_{j}}G_{J_{i}}^{(1)}+g_{2}\partial_{g_{2}}G_{J_{i}}^{(1)}\bigg],\\
\frac{dg_{2}}{d\ell}={}&-g_{2}\bigg[\sum_{j}J_{j}\partial_{J_{j}}G_{g_{2}}^{(1)}+g_{2}\partial_{g_{2}}G_{g_{2}}^{(1)}\bigg],
\end{align}
where we expand $G_{J_{j}}=Z_{f}^{-1}Z_{c}^{-1}Z_{J_{j}}$ and $G_{g_{2}}=Z_{c}^{-2}Z_{g_{2}}$ in order to get $G_{J_{j}}^{(1)}$ and $G_{g_{2}}^{(1)}$ as follows:
\begin{align}
G_{J_{j}}=Z_{f}^{-1}Z_{c}^{-1}Z_{J_{j}}=\sum_{m=0}^{\infty}\frac{G_{J_{j}}^{(m)}(\{J_{k},g_{2}\})}{\epsilon'^{m}},\\
G_{g_{2}}=Z_{c}^{-2}Z_{g_{2}}=\sum_{m=0}^{\infty}\frac{G_{g_{2}}^{(m)}(\{J_{k},g_{2}\})}{\epsilon'^{m}},
\end{align}
and the first terms $G_{J_{j}}^{(0)}=G_{g_{2}}^{(0)}=1$ and $i,j,k={{\perp}F},z F,{\perp}B,z B$.
After redefining the parameters, $\tilde{J}_{i}\equiv N_{0}J_{i}$ and $\tilde{g}_{2}\equiv N_{0}g_{2}$, the RG flow equations are
\begin{align}
\frac{d\tilde{J}_{\perp F}}{d\ell}={}&(\tilde{J}_{\perp F}\tilde{J}_{z F}+\tilde{J}_{\perp B}\tilde{J}_{z B})\notag\\&-\tilde{J}_{\perp F}(\tilde{J}_{\perp F}^{2}+\tilde{J}_{z F}^{2}+\tilde{J}_{\perp B}^{2}+\tilde{J}_{z B}^{2})/2,	\label{eq:JpFn2}\\
\frac{d\tilde{J}_{z F}}{d\ell}={}&(\tilde{J}_{\perp F}^{2}+\tilde{J}_{\perp B}^{2})-\tilde{J}_{z F}(\tilde{J}_{\perp F}^{2}+\tilde{J}_{\perp B}^{2}),\\
\frac{d\tilde{J}_{\perp B}}{d\ell}={}&\tilde{g}_{2}(1-\tilde{g}_{2})\tilde{J}_{\perp B}+(\tilde{J}_{\perp F}\tilde{J}_{z B}+\tilde{J}_{\perp B}\tilde{J}_{z F})\notag\\&-\tilde{J}_{\perp B}(\tilde{J}_{\perp F}^{2}+\tilde{J}_{z F}^{2}+\tilde{J}_{\perp B}^{2}+\tilde{J}_{z B}^{2})/2,\\
\frac{d\tilde{J}_{z B}}{d\ell}={}&\tilde{g}_{2}(1-\tilde{g}_{2})\tilde{J}_{z B}+2\tilde{J}_{\perp F}\tilde{J}_{\perp B}\notag\\&-\tilde{J}_{z B}(\tilde{J}_{\perp F}^{2}+\tilde{J}_{\perp B}^{2}),\\
\frac{d\tilde{g}_{2}}{d\ell}={}&0.	\label{eq:g2n2}
\end{align}
The fixed points of above RG flow equations are explained in the main text.
In the non-Hermitian case, the Kondo couplings are written as $\tilde{J}_{ab}=\tilde{J}_{ab1}+i\tilde{J}_{ab2}$ where $\tilde{J}_{ab1/ab2}$ are real-valued. The RG flow equations are
\begin{mywidetext}
\begin{align}
\frac{d\tilde{J}_{\perp F1}}{d\ell}={}&(\tilde{J}_{\perp F1}\tilde{J}_{z F1}-\tilde{J}_{\perp F2}\tilde{J}_{z F2}+\tilde{J}_{\perp B1}\tilde{J}_{z B1}-\tilde{J}_{\perp B2}\tilde{J}_{z B2})\notag\\
&-\tilde{J}_{\perp F1}(\tilde{J}_{\perp F1}^{2}+\tilde{J}_{z F1}^{2}+\tilde{J}_{\perp B1}^{2}+\tilde{J}_{z B1}^{2}-\tilde{J}_{\perp F2}^{2}-\tilde{J}_{z F2}^{2}-\tilde{J}_{\perp B2}^{2}-\tilde{J}_{z B2}^{2})/2\notag\\
&+\tilde{J}_{\perp F2}(\tilde{J}_{\perp F1}\tilde{J}_{\perp F2}+\tilde{J}_{z F1}\tilde{J}_{z F2}+\tilde{J}_{\perp B1}\tilde{J}_{\perp B2}+\tilde{J}_{z B1}\tilde{J}_{z B2}),\label{eq:JpF1}\\
\frac{d\tilde{J}_{\perp F2}}{d\ell}={}&(\tilde{J}_{\perp F2}\tilde{J}_{z F1}+\tilde{J}_{\perp F1}\tilde{J}_{z F2}+\tilde{J}_{\perp B2}\tilde{J}_{z B1}+\tilde{J}_{\perp B1}\tilde{J}_{z B2})\notag\\
&-\tilde{J}_{\perp F2}(\tilde{J}_{\perp F1}^{2}+\tilde{J}_{z F1}^{2}+\tilde{J}_{\perp B1}^{2}+\tilde{J}_{z B1}^{2}-\tilde{J}_{\perp F2}^{2}-\tilde{J}_{z F2}^{2}-\tilde{J}_{\perp B2}^{2}-\tilde{J}_{z B2}^{2})/2\notag\\
&-\tilde{J}_{\perp F1}(\tilde{J}_{\perp F1}\tilde{J}_{\perp F2}+\tilde{J}_{z F1}\tilde{J}_{z F2}+\tilde{J}_{\perp B1}\tilde{J}_{\perp B2}+\tilde{J}_{z B1}\tilde{J}_{z B2}),\\
\frac{d\tilde{J}_{z F1}}{d\ell}={}&(\tilde{J}_{\perp F1}^{2}+\tilde{J}_{\perp B1}^{2}-\tilde{J}_{\perp F2}^{2}-\tilde{J}_{\perp B2}^{2})\notag\\
&-\tilde{J}_{z F1}(\tilde{J}_{\perp F1}^{2}+\tilde{J}_{\perp B1}^{2}-\tilde{J}_{\perp F2}^{2}-\tilde{J}_{\perp B2}^{2})+2\tilde{J}_{z F2}(\tilde{J}_{\perp F1}\tilde{J}_{\perp F2}+\tilde{J}_{\perp B1}\tilde{J}_{\perp B2}),\\
\frac{d\tilde{J}_{z F2}}{d\ell}={}&2(\tilde{J}_{\perp F1}\tilde{J}_{\perp F2}+\tilde{J}_{\perp B1}\tilde{J}_{\perp B2})\notag\\
&-\tilde{J}_{z F2}(\tilde{J}_{\perp F1}^{2}+\tilde{J}_{\perp B1}^{2}-\tilde{J}_{\perp F2}^{2}-\tilde{J}_{\perp B2}^{2})-2\tilde{J}_{z F1}(\tilde{J}_{\perp F1}\tilde{J}_{\perp F2}+\tilde{J}_{\perp B1}\tilde{J}_{\perp B2}),\\
\frac{d\tilde{J}_{\perp B1}}{d\ell}={}&\tilde{g}_{2}(1-\tilde{g}_{2})\tilde{J}_{\perp B1}+(\tilde{J}_{\perp F1}\tilde{J}_{z B1}-\tilde{J}_{\perp F2}\tilde{J}_{z B2}+\tilde{J}_{\perp B1}\tilde{J}_{z F1}-\tilde{J}_{\perp B2}\tilde{J}_{z F2})\notag\\
&-\tilde{J}_{\perp B1}(\tilde{J}_{\perp F1}^{2}+\tilde{J}_{z F1}^{2}+\tilde{J}_{\perp B1}^{2}+\tilde{J}_{z B1}^{2}-\tilde{J}_{\perp F2}^{2}-\tilde{J}_{z F2}^{2}-\tilde{J}_{\perp B2}^{2}-\tilde{J}_{z B2}^{2})/2\notag\\
&+\tilde{J}_{\perp B2}(\tilde{J}_{\perp F1}\tilde{J}_{\perp F2}+\tilde{J}_{z F1}\tilde{J}_{z F2}+\tilde{J}_{\perp B1}\tilde{J}_{\perp B2}+\tilde{J}_{z B1}\tilde{J}_{z B2}),\\
\frac{d\tilde{J}_{\perp B2}}{d\ell}={}&\tilde{g}_{2}(1-\tilde{g}_{2})\tilde{J}_{\perp B2}+(\tilde{J}_{\perp F2}\tilde{J}_{z B1}+\tilde{J}_{\perp F1}\tilde{J}_{z B2}+\tilde{J}_{\perp B2}\tilde{J}_{z F1}+\tilde{J}_{\perp B1}\tilde{J}_{z F2})\notag\\
&-\tilde{J}_{\perp B2}(\tilde{J}_{\perp F1}^{2}+\tilde{J}_{z F1}^{2}+\tilde{J}_{\perp B1}^{2}+\tilde{J}_{z B1}^{2}-\tilde{J}_{\perp F2}^{2}-\tilde{J}_{z F2}^{2}-\tilde{J}_{\perp B2}^{2}-\tilde{J}_{z B2}^{2})/2\notag\\
&-\tilde{J}_{\perp B1}(\tilde{J}_{\perp F1}\tilde{J}_{\perp F2}+\tilde{J}_{z F1}\tilde{J}_{z F2}+\tilde{J}_{\perp B1}\tilde{J}_{\perp B2}+\tilde{J}_{z B1}\tilde{J}_{z B2}),\\
\frac{d\tilde{J}_{z B1}}{d\ell}={}&\tilde{g}_{2}(1-\tilde{g}_{2})\tilde{J}_{z B1}+2(\tilde{J}_{\perp F1}\tilde{J}_{\perp B1}-\tilde{J}_{\perp F2}\tilde{J}_{\perp B2})\notag\\
&-\tilde{J}_{z B1}(\tilde{J}_{\perp F1}^{2}+\tilde{J}_{\perp B1}^{2}-\tilde{J}_{\perp F2}^{2}-\tilde{J}_{\perp B2}^{2})+2\tilde{J}_{z B2}(\tilde{J}_{\perp F1}\tilde{J}_{\perp F2}+\tilde{J}_{\perp B1}\tilde{J}_{\perp B2}),\\
\frac{d\tilde{J}_{z B2}}{d\ell}={}&\tilde{g}_{2}(1-\tilde{g}_{2})\tilde{J}_{z B2}+2(\tilde{J}_{\perp F1}\tilde{J}_{\perp B2}+\tilde{J}_{\perp F2}\tilde{J}_{\perp B1})\notag\\
&-\tilde{J}_{z B2}(\tilde{J}_{\perp F1}^{2}+\tilde{J}_{\perp B1}^{2}-\tilde{J}_{\perp F2}^{2}-\tilde{J}_{\perp B2}^{2})-2\tilde{J}_{z B1}(\tilde{J}_{\perp F1}\tilde{J}_{\perp F2}+\tilde{J}_{\perp B1}\tilde{J}_{\perp B2}).\label{eq:JzB2}
\end{align}

At the isotropic SU(2) limit, $\tilde{J}_{\perp F}=\tilde{J}_{z F}=\tilde{J}_{F}$ and $\tilde{J}_{\perp B}=\tilde{J}_{z B}=\tilde{J}_{B}$, Eqs.~(\ref{eq:JpFn2})-(\ref{eq:g2n2}) can be reduced by
\begin{align}
\frac{d\tilde{J}_{F}}{d\ell}={}&(\tilde{J}_{F}^{2}+\tilde{J}_{B}^{2})-\tilde{J}_{F}(\tilde{J}_{F}^{2}+\tilde{J}_{B}^{2}),\\
\frac{d\tilde{J}_{B}}{d\ell}={}&\tilde{g}_{2}(1-\tilde{g}_{2})\tilde{J}_{B}+2\tilde{J}_{F}\tilde{J}_{B}-\tilde{J}_{B}(\tilde{J}_{F}^{2}+\tilde{J}_{B}^{2}),
\end{align}
where we omit $d\tilde{g}_{2}/d\ell$ because it is zero.
\end{mywidetext}
%
\end{document}